\documentclass[twocolumn,aps,showpacs,showkeys]{revtex4}  

\usepackage{graphicx}
\usepackage{longtable}

\begin{document}

\author{Richard J. Mathar}
\email{mathar@strw.leidenuniv.nl}
\homepage{http://www.strw.leidenuniv.nl/~mathar}
\affiliation{
Leiden Observatory, Leiden University, P.O. Box 9513, 2300 RA Leiden, The Netherlands}

\date{\today}
\title{Refractive Index of Humid Air in the Infrared: Model Fits}

\begin{abstract}
The theory of summation of electromagnetic line transitions
is used to tabulate the
Taylor expansion of the refractive index of humid air over the basic
independent parameters (temperature, pressure, humidity, wavelength) in five separate
infrared regions from the H to the Q band at a fixed percentage of Carbon Dioxide.
These are least-squares fits to raw, highly resolved spectra for a set of temperatures
from 10 to 25 $^\circ$C, a set of pressures from 500 to 1023 hPa, and a set of
relative humidities from 5 to 60\%. These choices reflect the prospective application to
characterize ambient air at mountain altitudes of astronomical telescopes.
\end{abstract}

\maketitle
\section{Scope} 
The paper provides easy access to predictions of the refractive index
of humid air at conditions that are typical in atmospheric physics, in 
support of ray tracing \cite{BertonJopt8} and astronomical applications
\cite{BasdenMNRAS357,ColavitaPASP116,MatharArxiv0605,MeisnerSPIE4838}
until experimental coverage of the infrared wavelengths
might render these obsolete.
The approach is in continuation of earlier work \cite{MatharAO43}
based on a more recent HITRAN database \cite{RothmanJQSRT96}
plus more precise
accounting of various electromagnetic effects for the dielectric response
of dilute gases, as described below.

The literature of optical, chemical and atmospheric physics on the
subject of the refractive index of moist air
falls into several categories, sorted with respect to 
decreasing relevance (if relevance is measured by the closeness to
experimental data and the degree of independence to the formalism
employed here):
\begin{enumerate}
\item
experiments on moist air in the visible
\cite{BarrellPTRS238,BonschMet35,NewboundJOSA39,CuthbertsonPTRS213},
\item
experiments on pure water vapor at 3.4 and 10.6 $\mu$m
\cite{Matsumoto82,Matsumoto84,MarchettiIPT48},
\item
experiments on dry air and its molecular constituents
in the visible \cite{BirchJOSA8,Hou,EdlenJOSA43,ZhangOL30},
at 1.064 $\mu$m \cite{VelskoAO25,PeckAO25},
up to 2.0 $\mu$m \cite{PeckJOSA52}, 
up to 1.7 $\mu$m \cite{PeckJOSA62,OldJOSA61,JhanwarCP67,RankJOSA48}, 
or at 10.6 $\mu$m \cite{SimmonsOptCo25,MarchettiIPT47},
\item
verification of the dispersion and higher order derivatives
with astronomical interferometry
\cite{TubbsSPIE5491},
\item
review formulas \cite{SchiebenerJPCRD19,Ciddor1996,Owens,JonesAO19},
\item
theoretical summation of electronic transitions \cite{MatharAO43,HillIP26,HillJOSA70,ColavitaPASP116}.
\end{enumerate}
The liquid and solid states of water are left aside, because extrapolation
of their dielectric response to the gaseous state is
difficult
by
the permanent dipole moment of the molecule.
Already in the Q band and then at sub-millimeter wavelengths
\cite{ZhenhuiJQSRT83,ManabeIJIMW6,RuegerFIG22}
and eventually in the static limit \cite{FernandezJPCR24}, the
refractive index plotted as a function of wavelength is more and more
structured by individual lines.
Since we will not present these functions at high resolution
but smooth fits within several bands  in the infrared, their
spiky appearance sets a natural limit to the far-IR wavelength regions
that our approach may
cover.

\section{Dielectric Model} 
\subsection{Methodology} 

The complex valued dielectric function $n(\omega)$ of air
\begin{equation}
n = \sqrt{1+\bar\chi} \label{eq.alledl}
\end{equation}
is constructed from molecular dynamical polarizabilities
\begin{eqnarray}
\chi_m(\omega)=2N_mc^2\sum_l\frac{S_{ml}}{\omega_{0ml}}
\bigg(
\frac{1}{\omega+\omega_{0ml}-i\Gamma_{ml}/2} && \nonumber\\
-\frac{1}{\omega-\omega_{0ml}-i\Gamma_{ml}/2}
\bigg). && \label{eq.oscl}
\end{eqnarray}
$N_m$ are molecular number densities, $S_{ml}$ are the
line intensities for the transitions enumerated by $l$.
$\omega_{0ml}$
are the transition angular frequencies, $\Gamma_{ml}$ the full
linewidths at half maximum.
$c$ is the velocity of light in vacuum,
and $i$ the imaginary unit. The
line shape (\ref{eq.oscl}) adheres to
the complex-conjugate symmetry $\chi_m(\omega)=\chi_m^*(-\omega)$,
as required for
functions which are real-valued in the time domain.
The sign convention of $\Gamma_{ml}$ merely reflects a sign
choice in the Fourier Transforms and carries no real significance;
a sign in the Kramers--Kronig formulas is bound to it.
The integrated imaginary part is \cite{HilbornAJP50}
\begin{equation}
\int_0^\infty \Im \chi_m(\tilde\nu)d\tilde\nu = N_m \sum_l \frac{S_{ml}}{k_{0ml}},
\end{equation}
where $\tilde\nu=k/(2\pi)=\omega/(2\pi c)=1/\lambda$ is the wavenumber.

\begin{figure}[hbt]
\includegraphics[width=0.5\textwidth]{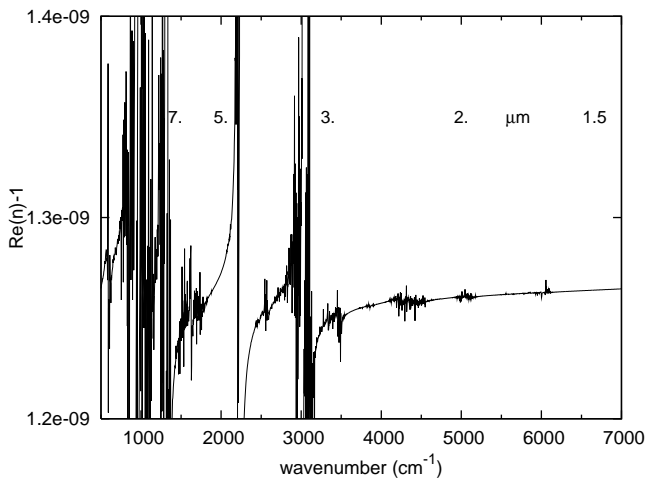}
\caption{
The combined refractivity of
0.39 Pa He,
0.075 Pa Kr,
0.038 Pa H$_2$,
0.030 Pa NH$_3$,
0.023 Pa N$_2$O,
and
$3.8\times 10^{-5}$ Pa SO$_2$,
is $\approx 10^{-9}$ at 12 $^\circ$C\@. The
dispersion (change in the refractivity) across the relevant bands
in the IR is $\approx 10^{-10}$.
}
\label{ParStd_nResid.ps}
\end{figure}

Line strengths
$S_{ml}$
and positions $\omega_{0ml}$
are based on the HITRAN \cite{RothmanJQSRT96} list
(see \cite{SmithJQSRT83,TanakaJMS239} for reviews)
and other sources as described earlier \cite{MatharAO43}.
The results of Section \ref{sec.resul}
include summation over the oscillator strengths of
the air components
N$_2$, O$_2$, Ar, Ne, CO$_2$, H$_2$O, O$_3$, CH$_4$ and CO\@.
Fig.\ \ref{ParStd_nResid.ps} is an estimate of the combined contribution
of tracer gases that are missing in this mix of molecules---and
effectively replaced by the average refractivity of the major components---
sorted with respect to abundance under rather clean
environmental conditions.
Their electromagnetic line lists are taken from
\cite{Ligten,YanAJ496} for He,
\cite{Chan2} for Kr,
\cite{Margoliash} for H$_2$,
\cite{RothmanJQSRT96,Margoliash} for N$_2$O,
\cite{RothmanJQSRT96} for SO$_2$,
and
\cite{RothmanJQSRT96,Margoliash} for NH$_3$.

The remaining subsections describe some refinements of the computer
program relative to its status three years ago \cite{MatharAO43}.

\subsection{Deviations from Ideal Gas Behavior}\label{Sec.virial} 

The second virial coefficients for Nitrogen and Oxygen are negative
at typical environmental temperatures, 
$\approx-7.5\times 10^{-6}$ m$^3$/mole for Nitrogen and $\approx -1.9\times 10^{-5}$ m$^3$/mole
for Oxygen at 12 $^\circ$C \cite{Dymond,WeberJResNatB74}.
More
molecules
are packed into a given volume at a given partial pressure than the
ideal gas equation predicts. The gas densities of Nitrogen and Oxygen
are $\approx 25$ mole/m$^3$ and $\approx 6$ mole/m$^3$ for Nitrogen and Oxygen, respectively,
at air pressures of the order 740 hPa, so
the correction factors for the density and for the refractivity
are of the order $2\times 10^{-4}$ due to this effect, the product of the
virial and the density. See \cite{LemmonJPCR29} for a review
and \cite{AchtermannJCP94} for examples of the crosstalk to refractivities.

The second and third virials for water vapor are larger
\cite{KellPRSA425,KusalikJCP103,HillIndECH27}---
the second $\approx -1.3\times 10^{-3}$ m$^3$/mole \cite[Fig 7.19]{WagnerJPCR31}---
and often placed as ``enhancement factors.''
See \cite{SatoJPCRF20} for a review on this subject and \cite{HarveyJPCR33}
for recent values of the second virial coefficient.
The values presented here use the equation of state in a self-consistent solution
of the first line of \cite[Tab.\ 6.3]{WagnerJPCR31} for water,
the values of \cite[p. 239]{Dymond} for Nitrogen,
and the NIST values \cite{NIST134}
for the second and third virials of Oxygen, Argon and Carbon Dioxide.

\subsection{Temperature Dependent Partition Sums} 
The temperature dependence of the partition sums
leads to temperature-dependent line strengths \cite{BarberMNRAS368}.
For the HITRAN lines, this has been implemented on a line-per-line basis
\cite{SimeckovaJQSRT98,GamacheJMolS517}.
The combined change induced by the upgrade to the
database of August '06
plus this increase of the line strengths at lower temperatures is
minuscule, less than $2\times 10^{-9}$ in the $c_{0\mathrm{ref}}$ coefficients
and less than $3\times 10^{-5}$ K in the $c_{0T}$ coefficients
reported in Section \ref{sec.resul}.

The line broadening parameters $\Gamma$ were not
changed \cite{TothJQSRT101,JacquemartJQSRT96}
from the ones at the HITRAN
reference pressure of 1 atm, since the effect on the
real part of the susceptibility is presumably negligible.
Effects of molecular clustering \cite{SlaninaJQSRT97} have not been
considered.

\subsection{Magnetic Susceptibility}
The paramagnetic susceptibility of Oxygen and diamagnetic
contribution of Nitrogen \cite{HavensPR43} account for most of the remaining
gap between 
theory and experiment. The volume susceptibility of dry air
is $\approx 3.7\times 10^{-7}$ \cite{DavisMet35} at 1013 hPa,
to increase $n$ by $\approx 1.3\times 10^{-7}$.
The magnetic dipole transitions of Oxygen \cite{BoreikoJQSRT32,KrupenieJPCR1}
are incorporated in the HITRAN list \cite{ChanceIJIMW12}, which
allows us to add dispersion to their response. (Since we are only dealing with the
limit of
small susceptibilities, the electric and magnetic susceptibilities
are additive, which means cross product terms have been neglected here.)
The magnetism of water is negligible because the magnetic moment
of the water molecule is
close to
the magnetic moment of
the Nitrogen molecule \cite{DavisMet35,CiniJCP49}, but the abundance
of the water molecules in air
much smaller than the abundance of Nitrogen molecules.

\subsection{Lorentz-Lorenz Cross Polarization}\label{Sec.LL}

We incorporate the mutual inter-molecular cross-polarization
with the Clausius-Mossotti (Lorentz-Lorenz) formula \cite{WolfJOSAA10,deGoedePhys58}:
the
macroscopic susceptibility $\bar \chi=n^2-1$ in Eq.\ (\ref{eq.alledl}).
is
\begin{equation}
\bar \chi = \frac{\sum \chi_m}{1-\sum\chi_m/3},
\label{eq.Lore2}
\end{equation}
where $\sum \chi_m$ is the sum over all atomic/molecular polarizabilities.
This increases the real part of $\chi$ by $\approx(\sum\chi_m)^2/3$,
hence the real part of $n$ by $\approx (\sum\chi_m)^2/6$,
which is of the order $2\times 10^{-8}$ if we take
$\sum_m \chi\approx 4\times 10^{-4}$ as a guideline.

\section{Comparison with Experiments} 

\begin{figure}[hbt]
\includegraphics[width=0.5\textwidth]{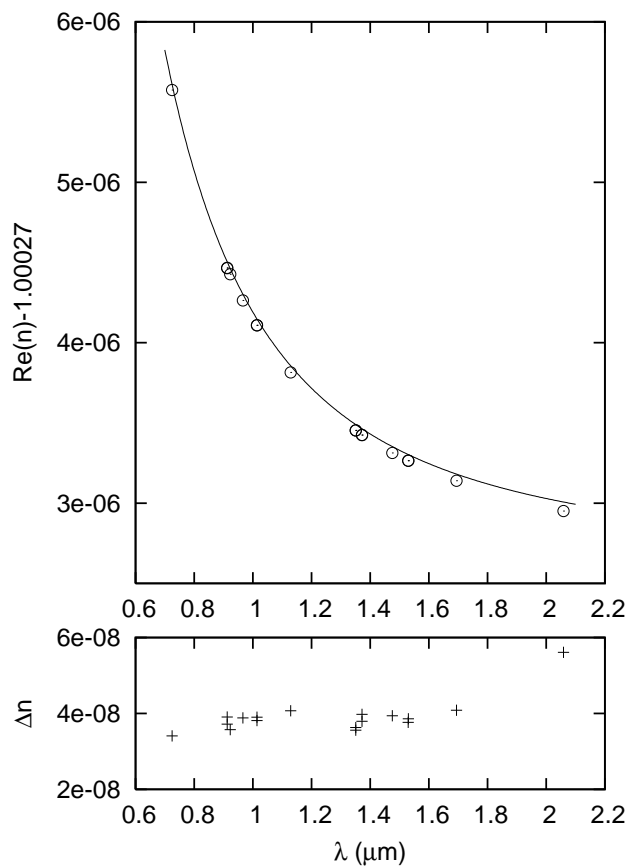}
\caption{
Top: The line are the raw data of the theory for dry air at 15 $^\circ$C
and 101324 Pa. Circles are the experimental values of Table I and II
by Peck and Reeder \cite{PeckJOSA62} plus one datum at 2.06 $\mu$m
by Peck and Khanna \cite{PeckJOSA52}. Bottom: Refractive index of the
theory minus these experimental values.
}
\label{PeckReeder.ps}
\end{figure}

The raw theoretical data (prior to the fit) exceeds experiments for dry
air in the visible and near infrared by $\approx 4\times 10^{-8}$
(Fig.\ \ref{PeckReeder.ps}).
Roughly $0.8\times 10^{-8}$ of this can be attributed to a change
in temperature scales \cite{BirchMet30}, and roughly $1\times 10^{-8}$
to a presumably lower CO$_2$ contents of 300 ppmv in \cite{PeckJOSA62}.

The pressure coefficient $c_{0p}$ for dry air at 10 $\mu$m in Table \ref{tab.fitn_N}
is to be compared to the value of $2.668\times 10^{-4}$/atm$=0.2633\times 10^{-8}$/Pa 
measured at $\lambda=10.57$ $\mu$m and $T=23$ $^\circ$C by Marchetti
and Simili \cite[Tab.\ 1]{MarchettiIPT47}.
More accurately, the pressure gradient predicted from
(\ref{eq.fit}) is
\begin{eqnarray}
\frac{\partial n}{\partial p} = \sum_{i=0,1,\ldots} &&
\Big[
c_{ip}
+ 2c_{ipp}\left(p-p_\mathrm{ref} \right)
+ c_{iTp}\left( \frac{1}{T}-\frac{1}{T_\mathrm{ref}}\right)
\nonumber \\
&&
+ c_{iHp}\left( H-H_\mathrm{ref} \right)
\Big]
\left(\tilde\nu - \tilde\nu_\mathrm{ref}\right)^i
\end{eqnarray}
and generates 
$0.2618\times 10^{-8}$/Pa
at the same wavelength, the same temperature, and
a  pressure---undocumented by Marchetti and Simili---of 1013.25 hPa.
The relative deviation of $6\times 10^{-3}$
between experiment and theory is still compatible with the error $5\times 10^{-3}$
provided by Marchetti and Simili.

\begin{figure}[hbt]
\includegraphics[width=0.5\textwidth]{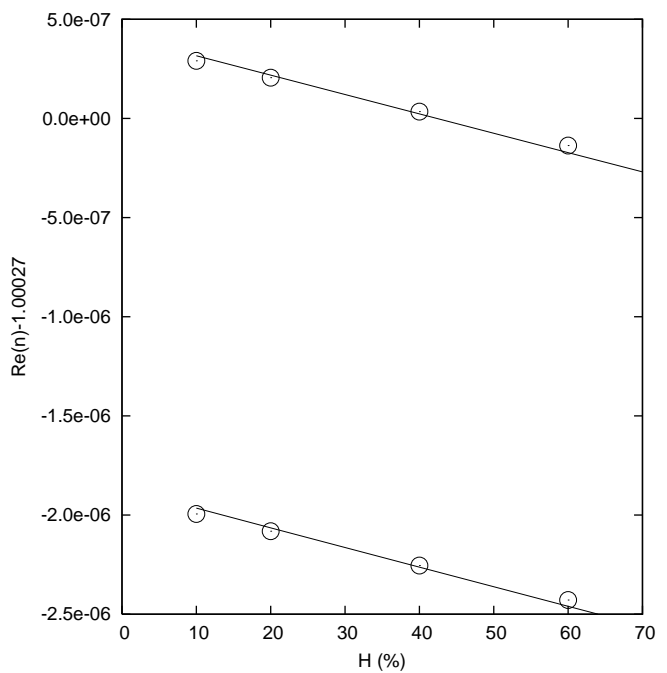}
\caption{
Lines are the raw data of the theory for humid air at 20 $^\circ$C
and 1000 hPa at four levels of humidity. Circles are from the
B\"onsch-Potulski fitting equation (at 370 ppm CO$_2$)
to their experimental data \cite{BonschMet35}.
The two groups of four comparisons refer to the
wavelengths
0.5087 and 0.644 $\mu$m.
}
\label{Bonsch.ps}
\end{figure}

The theory deviates from the humid air data at the longest two wavelengths
of the B\"onsch-Potulski experiments \cite{BonschMet35} by
$3.5\times 10^{-8}$ or less: Fig.\ \ref{Bonsch.ps}.

\begin{figure}[hbt]
\includegraphics[width=0.5\textwidth]{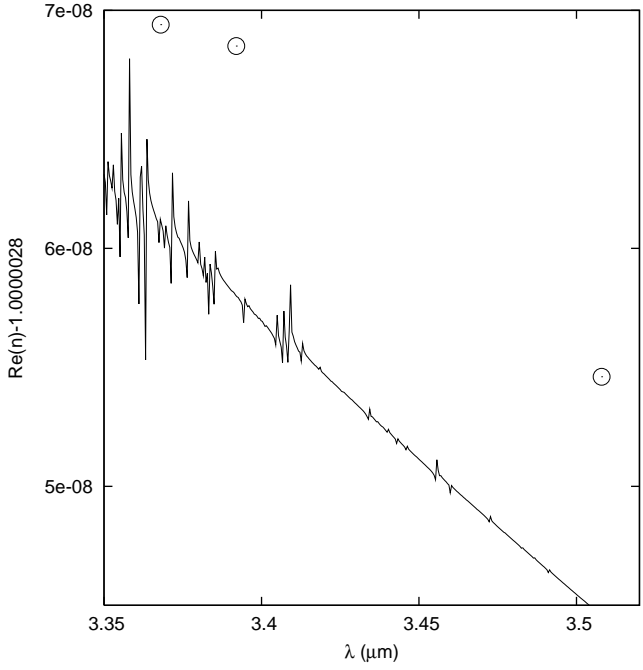}
\caption{Comparison of the theory (solid line) with the
experimental data by Matsumoto (3 circles) \cite[Tab.\ 1]{Matsumoto82} for water vapor
of 1333 Pa
at 20 $^\circ$C.
}
\label{MatsuH2O3.ps}
\end{figure}

\begin{figure}[hbt]
\includegraphics[width=0.5\textwidth]{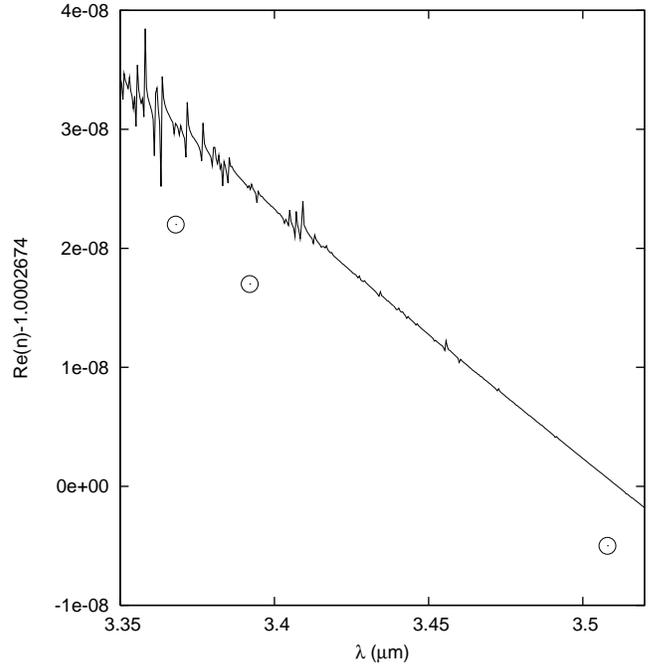}
\caption{Comparison of the theory (solid line, 300 ppmv CO$_2$) with
the experimental data by Matsumoto (3 circles) \cite[Tab.\ 2]{Matsumoto82} for humid air
at $p=1013.25$ hPa, $T=20$ $^\circ$C and $H=56.98$ \%.
}
\label{Matsudry3.ps}
\end{figure}

The difference between the theory and experiments with pure water
vapor (Fig.\ \ref{MatsuH2O3.ps}) and moist air (Fig.\ \ref{Matsudry3.ps})
is $\approx 1\times 10^{-8}$ at 3.4 $\mu$m.

\section{Results}\label{sec.resul} 

The bold least squares fit to the raw data---examples of which are shown
in \cite[Figs.\ 2,12]{MatharArxiv0605}---looks as follows:
\begin{eqnarray}
n-1 &=& \sum_{i=0,1,2,\ldots} c_i(T,p,H) \left(\tilde\nu - \tilde\nu_\mathrm{ref}\right)^i ;
\\
c_i(T,p,H) &=& c_{i\mathrm{ref}}
\nonumber \\ &&
+ c_{iT}\left( \frac{1}{T}-\frac{1}{T_\mathrm{ref}}\right)
+ c_{iTT}\left( \frac{1}{T}-\frac{1}{T_\mathrm{ref}}\right)^2
\nonumber \\ &&
+ c_{iH}\left(H-H_\mathrm{ref} \right)
+ c_{iHH}\left(H-H_\mathrm{ref} \right)^2
\nonumber \\ &&
+ c_{ip}\left(p-p_\mathrm{ref} \right)
+ c_{ipp}\left(p-p_\mathrm{ref} \right)^2
\nonumber \\ &&
+ c_{iTH}\left( \frac{1}{T}-\frac{1}{T_\mathrm{ref}}\right)
         \left(H-H_\mathrm{ref} \right)
\nonumber \\ &&
 + c_{iTp}\left( \frac{1}{T}-\frac{1}{T_\mathrm{ref}}\right)
          \left(p-p_\mathrm{ref} \right)
\nonumber \\ &&
+ c_{iHp}\left( H-H_\mathrm{ref} \right)
         \left(p-p_\mathrm{ref} \right).
\label{eq.fit}
\end{eqnarray}

\begin{table*}[hbt]
\caption{
Fitting coefficients for the multivariate Taylor expansion (\ref{eq.fit})
to the real part of the index of refraction
over the $1.3\le 1/\tilde\nu\le 2.5$ $\mu$m range
with $\tilde\nu_\mathrm{ref}=10^4/2.25$ cm$^{-1}$.
}
\label{tab.fitn_K}
\begin{ruledtabular}
\begin{tabular}{c|ccccc}
$i$ & $c_{i\mathrm{ref}}$ / cm$^i$ & $c_{iT}$ / cm$^i$K & $c_{iTT}$ / [cm$^i$K$^2$] &
  $c_{iH}$ / [cm$^i$/\%] & $c_{iHH}$ / [cm$^i$/\%$^2$]
   \\
\hline
0 & $ 0.200192\times 10^{-3}$
& $ 0.588625\times 10^{-1}$
& $-3.01513$
& $-0.103945\times 10^{-7}$
& $ 0.573256\times 10^{-12}$ \\

1& $ 0.113474\times 10^{-9}$
& $-0.385766\times 10^{-7}$
& $ 0.406167\times 10^{-3}$
& $ 0.136858\times 10^{-11}$
& $ 0.186367\times 10^{-16}$ \\

2& $-0.424595\times 10^{-14}$
& $ 0.888019\times 10^{-10}$
& $-0.514544\times 10^{-6}$
& $-0.171039\times 10^{-14}$
& $-0.228150\times 10^{-19}$ \\

3& $ 0.100957\times 10^{-16}$
& $-0.567650\times 10^{-13}$
& $ 0.343161\times 10^{-9}$
& $ 0.112908\times 10^{-17}$
& $ 0.150947\times 10^{-22}$ \\

4& $-0.293315\times 10^{-20}$
& $ 0.166615\times 10^{-16}$
& $-0.101189\times 10^{-12}$
& $-0.329925\times 10^{-21}$
& $-0.441214\times 10^{-26}$ \\

5& $ 0.307228\times 10^{-24}$
& $-0.174845\times 10^{-20}$
& $ 0.106749\times 10^{-16}$
& $ 0.344747\times 10^{-25}$
& $ 0.461209\times 10^{-30}$ \\
\end{tabular}

\begin{tabular}{c|ccccc}
$i$ & $c_{ip}$ / [cm$^i$/Pa] & $c_{ipp}$ / [cm$^i$/Pa$^2$] & $c_{iTH}$  / [cm$^i$K/\%]
   & $c_{iTp}$ / [cm$^i$K/Pa] & $c_{iHp}$ / [cm$^i$/(\% Pa)] \\
\hline
0 & $ 0.267085\times 10^{-8}$
& $ 0.609186\times 10^{-17}$
& $ 0.497859\times 10^{-4}$
& $ 0.779176\times 10^{-6}$
& $-0.206567\times 10^{-15}$ \\
1& $ 0.135941\times 10^{-14}$
& $ 0.519024\times 10^{-23}$
& $-0.661752\times 10^{-8}$
& $ 0.396499\times 10^{-12}$
& $ 0.106141\times 10^{-20}$ \\
2& $ 0.135295\times 10^{-18}$
& $-0.419477\times 10^{-27}$
& $ 0.832034\times 10^{-11}$
& $ 0.395114\times 10^{-16}$
& $-0.149982\times 10^{-23}$ \\
3& $ 0.818218\times 10^{-23}$
& $ 0.434120\times 10^{-30}$
& $-0.551793\times 10^{-14}$
& $ 0.233587\times 10^{-20}$
& $ 0.984046\times 10^{-27}$ \\
4& $-0.222957\times 10^{-26}$
& $-0.122445\times 10^{-33}$
& $ 0.161899\times 10^{-17}$
& $-0.636441\times 10^{-24}$
& $-0.288266\times 10^{-30}$ \\
5& $ 0.249964\times 10^{-30}$
& $ 0.134816\times 10^{-37}$
& $-0.169901\times 10^{-21}$
& $ 0.716868\times 10^{-28}$
& $ 0.299105\times 10^{-34}$ \\
\end{tabular}
\end{ruledtabular}
\end{table*}

\begin{table*}[hbt]
\caption{
Fitting coefficients for the multivariate Taylor expansion (\ref{eq.fit})
to the real part of the index of refraction
over the $2.8\le 1/\tilde\nu\le 4.2$ $\mu$m range
with $\tilde\nu_\mathrm{ref}=10^4/3.4$ cm$^{-1}$.
}
\label{tab.fitn_L}
\begin{ruledtabular}
\begin{tabular}{c|ccccc}
$i$ & $c_{i\mathrm{ref}}$ / cm$^i$ & $c_{iT}$ / cm$^i$K & $c_{iTT}$ / [cm$^i$K$^2$] &
  $c_{iH}$ / [cm$^i$/\%] & $c_{iHH}$ / [cm$^i$/\%$^2$]
   \\
\hline
0& $ 0.200049\times 10^{-3}$
 & $ 0.588431\times 10^{-1}$
 & $-3.13579$
 & $-0.108142\times 10^{-7}$
 & $ 0.586812\times 10^{-12}$ \\

1& $ 0.145221\times 10^{-9}$
 & $-0.825182\times 10^{-7}$
 & $ 0.694124\times 10^{-3}$
 & $ 0.230102\times 10^{-11}$
 & $ 0.312198\times 10^{-16}$ \\

2& $ 0.250951\times 10^{-12}$
 & $ 0.137982\times 10^{-9}$
 & $-0.500604\times 10^{-6}$
 & $-0.154652\times 10^{-14}$
 & $-0.197792\times 10^{-19}$ \\

3& $-0.745834\times 10^{-15}$
 & $ 0.352420\times 10^{-13}$
 & $-0.116668\times 10^{-8}$
 & $-0.323014\times 10^{-17}$
 & $-0.461945\times 10^{-22}$ \\

4& $-0.161432\times 10^{-17}$
 & $-0.730651\times 10^{-15}$
 & $ 0.209644\times 10^{-11}$
 & $ 0.630616\times 10^{-20}$
 & $ 0.788398\times 10^{-25}$ \\

5& $ 0.352780\times 10^{-20}$
 & $-0.167911\times 10^{-18}$
 & $ 0.591037\times 10^{-14}$
 & $ 0.173880\times 10^{-22}$
 & $ 0.245580\times 10^{-27}$ \\
\end{tabular}

\begin{tabular}{c|ccccc}
$i$ & $c_{ip}$ / [cm$^i$/Pa] & $c_{ipp}$ / [cm$^i$/Pa$^2$] & $c_{iTH}$  / [cm$^i$K/\%]
   & $c_{iTp}$ / [cm$^i$K/Pa] & $c_{iHp}$ / [cm$^i$/(\% Pa)] \\
\hline
0& $ 0.266900\times 10^{-8}$
 & $ 0.608860\times 10^{-17}$
 & $ 0.517962\times 10^{-4}$
 & $ 0.778638\times 10^{-6}$
 & $-0.217243\times 10^{-15}$ \\
1& $ 0.168162\times 10^{-14}$
 & $ 0.461560\times 10^{-22}$
 & $-0.112149\times 10^{-7}$
 & $ 0.446396\times 10^{-12}$
 & $ 0.104747\times 10^{-20}$ \\
2& $ 0.353075\times 10^{-17}$
 & $ 0.184282\times 10^{-24}$
 & $ 0.776507\times 10^{-11}$
 & $ 0.784600\times 10^{-15}$
 & $-0.523689\times 10^{-23}$ \\
3& $-0.963455\times 10^{-20}$
 & $-0.524471\times 10^{-27}$
 & $ 0.172569\times 10^{-13}$
 & $-0.195151\times 10^{-17}$
 & $ 0.817386\times 10^{-26}$ \\
4& $-0.223079\times 10^{-22}$
 & $-0.121299\times 10^{-29}$
 & $-0.320582\times 10^{-16}$
 & $-0.542083\times 10^{-20}$
 & $ 0.309913\times 10^{-28}$ \\
5& $ 0.453166\times 10^{-25}$
 & $ 0.246512\times 10^{-32}$
 & $-0.899435\times 10^{-19}$
 & $ 0.103530\times 10^{-22}$
 & $-0.363491\times 10^{-31}$ \\
\end{tabular}
\end{ruledtabular}
\end{table*}

\begin{table*}[hbt]
\caption{
Fitting coefficients for the multivariate Taylor expansion (\ref{eq.fit})
to the real part of the index of refraction
over the $4.35\le 1/\tilde\nu\le 5.3$ $\mu$m range
with $\tilde\nu_\mathrm{ref}=10^4/4.8$ cm$^{-1}$.
}
\label{tab.fitn_M}
\begin{ruledtabular}
\begin{tabular}{c|ccccc}
$i$ & $c_{i\mathrm{ref}}$ / cm$^i$ & $c_{iT}$ / cm$^i$K & $c_{iTT}$ / [cm$^i$K$^2$] &
  $c_{iH}$ / [cm$^i$/\%] & $c_{iHH}$ / [cm$^i$/\%$^2$]
   \\
\hline
0& $ 0.200020\times 10^{-3}$
 & $ 0.590035\times 10^{-1}$
 & $-4.09830$
 & $-0.140463\times 10^{-7}$
 & $ 0.543605\times 10^{-12}$ \\

1& $ 0.275346\times 10^{-9}$
 & $-0.375764\times 10^{-6}$
 & $ 0.250037\times 10^{-2}$
 & $ 0.839350\times 10^{-11}$
 & $ 0.112802\times 10^{-15}$ \\

2& $ 0.325702\times 10^{-12}$
 & $ 0.134585\times 10^{-9}$
 & $ 0.275187\times 10^{-6}$
 & $-0.190929\times 10^{-14}$
 & $-0.229979\times 10^{-19}$ \\

3& $-0.693603\times 10^{-14}$
 & $ 0.124316\times 10^{-11}$
 & $-0.653398\times 10^{-8}$
 & $-0.121399\times 10^{-16}$
 & $-0.191450\times 10^{-21}$ \\

4& $ 0.285610\times 10^{-17}$
 & $ 0.508510\times 10^{-13}$
 & $-0.310589\times 10^{-9}$
 & $-0.898863\times 10^{-18}$
 & $-0.120352\times 10^{-22}$ \\

5& $ 0.338758\times 10^{-18}$
 & $-0.189245\times 10^{-15}$
 & $ 0.127747\times 10^{-11}$
 & $ 0.364662\times 10^{-20}$
 & $ 0.500955\times 10^{-25}$ \\
\end{tabular}

\begin{tabular}{c|ccccc}
$i$ & $c_{ip}$ / [cm$^i$/Pa] & $c_{ipp}$ / [cm$^i$/Pa$^2$] & $c_{iTH}$  / [cm$^i$K/\%]
   & $c_{iTp}$ / [cm$^i$K/Pa] & $c_{iHp}$ / [cm$^i$/(\% Pa)] \\
\hline
0& $ 0.266898\times 10^{-8}$
 & $ 0.610706\times 10^{-17}$
 & $ 0.674488\times 10^{-4}$
 & $ 0.778627\times 10^{-6}$
 & $-0.211676\times 10^{-15}$ \\
1& $ 0.273629\times 10^{-14}$
 & $ 0.116620\times 10^{-21}$
 & $-0.406775\times 10^{-7}$
 & $ 0.593296\times 10^{-12}$
 & $ 0.487921\times 10^{-20}$ \\
2& $ 0.463466\times 10^{-17}$
 & $ 0.244736\times 10^{-24}$
 & $ 0.289063\times 10^{-11}$
 & $ 0.145042\times 10^{-14}$
 & $-0.682545\times 10^{-23}$ \\
3& $-0.916894\times 10^{-19}$
 & $-0.497682\times 10^{-26}$
 & $ 0.819898\times 10^{-13}$
 & $ 0.489815\times 10^{-17}$
 & $ 0.942802\times 10^{-25}$ \\
4& $ 0.136685\times 10^{-21}$
 & $ 0.742024\times 10^{-29}$
 & $ 0.468386\times 10^{-14}$
 & $ 0.327941\times 10^{-19}$
 & $-0.946422\times 10^{-27}$ \\
5& $ 0.413687\times 10^{-23}$
 & $ 0.224625\times 10^{-30}$
 & $-0.191182\times 10^{-16}$
 & $ 0.128020\times 10^{-21}$
 & $-0.153682\times 10^{-29}$ \\
\end{tabular}
\end{ruledtabular}
\end{table*}

\begin{table*}[hbt]
\caption{
Fitting coefficients for the multivariate Taylor expansion (\ref{eq.fit})
to the real part of the index of refraction
over the $7.5\le 1/\tilde\nu\le 14.1$ $\mu$m range
with $\tilde\nu_\mathrm{ref}=10^4/10.1$ cm$^{-1}$.
}
\label{tab.fitn_N}
\begin{ruledtabular}
\begin{tabular}{c|ccccc}
$i$ & $c_{i\mathrm{ref}}$ / cm$^i$ & $c_{iT}$ / cm$^i$K & $c_{iTT}$ / [cm$^i$K$^2$] &
  $c_{iH}$ / [cm$^i$/\%] & $c_{iHH}$ / [cm$^i$/\%$^2$]
   \\
\hline
0& $ 0.199885\times 10^{-3}$
 & $ 0.593900\times 10^{-1}$
 & $-6.50355$
 & $-0.221938\times 10^{-7}$
 & $ 0.393524\times 10^{-12}$ \\

1& $ 0.344739\times 10^{-9}$
 & $-0.172226\times 10^{-5}$
 & $ 0.103830\times 10^{-1}$
 & $ 0.347377\times 10^{-10}$
 & $ 0.464083\times 10^{-15}$ \\

2& $-0.273714\times 10^{-12}$
 & $ 0.237654\times 10^{-8}$
 & $-0.139464\times 10^{-4}$
 & $-0.465991\times 10^{-13}$
 & $-0.621764\times 10^{-18}$ \\

3& $ 0.393383\times 10^{-15}$
 & $-0.381812\times 10^{-11}$
 & $ 0.220077\times 10^{-7}$
 & $ 0.735848\times 10^{-16}$
 & $ 0.981126\times 10^{-21}$ \\

4& $-0.569488\times 10^{-17}$
 & $ 0.305050\times 10^{-14}$
 & $-0.272412\times 10^{-10}$
 & $-0.897119\times 10^{-19}$
 & $-0.121384\times 10^{-23}$ \\

5& $ 0.164556\times 10^{-19}$
 & $-0.157464\times 10^{-16}$
 & $ 0.126364\times 10^{-12}$
 & $ 0.380817\times 10^{-21}$
 & $ 0.515111\times 10^{-26}$ \\
\end{tabular}

\begin{tabular}{c|ccccc}
$i$ & $c_{ip}$ / [cm$^i$/Pa] & $c_{ipp}$ / [cm$^i$/Pa$^2$] & $c_{iTH}$  / [cm$^i$K/\%]
   & $c_{iTp}$ / [cm$^i$K/Pa] & $c_{iHp}$ / [cm$^i$/(\% Pa)] \\
\hline
0& $ 0.266809\times 10^{-8}$
 & $ 0.610508\times 10^{-17}$
 & $ 0.106776\times 10^{-3}$
 & $ 0.778368\times 10^{-6}$
 & $-0.206365\times 10^{-15}$ \\
1& $ 0.695247\times 10^{-15}$
 & $ 0.227694\times 10^{-22}$
 & $-0.168516\times 10^{-6}$
 & $ 0.216404\times 10^{-12}$
 & $ 0.300234\times 10^{-19}$ \\
2& $ 0.159070\times 10^{-17}$
 & $ 0.786323\times 10^{-25}$
 & $ 0.226201\times 10^{-9}$
 & $ 0.581805\times 10^{-15}$
 & $-0.426519\times 10^{-22}$ \\
3& $-0.303451\times 10^{-20}$
 & $-0.174448\times 10^{-27}$
 & $-0.356457\times 10^{-12}$
 & $-0.189618\times 10^{-17}$
 & $ 0.684306\times 10^{-25}$ \\
4& $-0.661489\times 10^{-22}$
 & $-0.359791\times 10^{-29}$
 & $ 0.437980\times 10^{-15}$
 & $-0.198869\times 10^{-19}$
 & $-0.467320\times 10^{-29}$ \\
5& $ 0.178226\times 10^{-24}$
 & $ 0.978307\times 10^{-32}$
 & $-0.194545\times 10^{-17}$
 & $ 0.589381\times 10^{-22}$
 & $ 0.126117\times 10^{-30}$ \\
\end{tabular}
\end{ruledtabular}
\end{table*}

\begin{table*}[hbt]
\caption{
Fitting coefficients for the multivariate Taylor expansion (\ref{eq.fit})
to the real part of the index of refraction
over the $16\le 1/\tilde\nu\le 28$ $\mu$m range
with $\tilde\nu_\mathrm{ref}=10^4/20$ cm$^{-1}$.
}
\label{tab.fitn_Q}
\begin{ruledtabular}
\begin{tabular}{c|ccccc}
$i$ & $c_{i\mathrm{ref}}$ / cm$^i$ & $c_{iT}$ / cm$^i$K & $c_{iTT}$ / [cm$^i$K$^2$] &
  $c_{iH}$ / [cm$^i$/\%] & $c_{iHH}$ / [cm$^i$/\%$^2$]
   \\
\hline
0& $ 0.199436\times 10^{-3}$
 & $ 0.621723\times 10^{-1}$
 & $-23.2409$
 & $-0.772707\times 10^{-7}$
 & $-0.326604\times 10^{-12}$ \\

1& $ 0.299123\times 10^{-8}$
 & $-0.177074\times 10^{-4}$
 & $ 0.108557$
 & $ 0.347237\times 10^{-9}$
 & $ 0.463606\times 10^{-14}$ \\

2& $-0.214862\times 10^{-10}$
 & $ 0.152213\times 10^{-6}$
 & $-0.102439\times 10^{-2}$
 & $-0.272675\times 10^{-11}$
 & $-0.364272\times 10^{-16}$ \\

3& $ 0.143338\times 10^{-12}$
 & $-0.954584\times 10^{-9}$
 & $ 0.634072\times 10^{-5}$
 & $ 0.170858\times 10^{-13}$
 & $ 0.228756\times 10^{-18}$ \\

4& $ 0.122398\times 10^{-14}$
 & $-0.996706\times 10^{-11}$
 & $ 0.762517\times 10^{-7}$
 & $ 0.156889\times 10^{-15}$
 & $ 0.209502\times 10^{-20}$ \\

5& $-0.114628\times 10^{-16}$
 & $ 0.921476\times 10^{-13}$
 & $-0.675587\times 10^{-9}$
 & $-0.150004\times 10^{-17}$
 & $-0.200547\times 10^{-22}$ \\
\end{tabular}

\begin{tabular}{c|ccccc}
$i$ & $c_{ip}$ / [cm$^i$/Pa] & $c_{ipp}$ / [cm$^i$/Pa$^2$] & $c_{iTH}$  / [cm$^i$K/\%]
   & $c_{iTp}$ / [cm$^i$K/Pa] & $c_{iHp}$ / [cm$^i$/(\% Pa)] \\
\hline
0& $ 0.266827\times 10^{-8}$
 & $ 0.613675\times 10^{-17}$
 & $ 0.375974\times 10^{-3}$
 & $ 0.778436\times 10^{-6}$
 & $-0.272614\times 10^{-15}$ \\
1& $ 0.120788\times 10^{-14}$
 & $ 0.585494\times 10^{-22}$
 & $-0.171849\times 10^{-5}$
 & $ 0.461840\times 10^{-12}$
 & $ 0.304662\times 10^{-18}$ \\
2& $ 0.522646\times 10^{-17}$
 & $ 0.286055\times 10^{-24}$
 & $ 0.146704\times 10^{-7}$
 & $ 0.306229\times 10^{-14}$
 & $-0.239590\times 10^{-20}$ \\
3& $ 0.783027\times 10^{-19}$
 & $ 0.425193\times 10^{-26}$
 & $-0.917231\times 10^{-10}$
 & $-0.623183\times 10^{-16}$
 & $ 0.149285\times 10^{-22}$ \\
4& $ 0.753235\times 10^{-21}$
 & $ 0.413455\times 10^{-28}$
 & $-0.955922\times 10^{-12}$
 & $-0.161119\times 10^{-18}$
 & $ 0.136086\times 10^{-24}$ \\
5& $-0.228819\times 10^{-24}$
 & $-0.812941\times 10^{-32}$
 & $ 0.880502\times 10^{-14}$
 & $ 0.800756\times 10^{-20}$
 & $-0.130999\times 10^{-26}$ \\
\end{tabular}
\end{ruledtabular}
\end{table*}

Here, $T$ is the absolute temperature with a reference value of
$T_\mathrm{ref}=(273.15+17.5)$ K,
$p$ is the air pressure with a reference value set
at $p_\mathrm{ref}=75000$ Pa, $H$ the relative humidity between 0 and 100 with a
reference value set at $H_\mathrm{ref}=10$ \%, and $\tilde\nu$ the wavenumber
$1/\lambda$ with
a reference value set at $\tilde\nu_\mathrm{ref}$.
The
units of these reference values match those of the tabulated
coefficients.
The range 1.3--2.5 $\mu$m is covered by Table \ref{tab.fitn_K},
the range 2.8--4.2 $\mu$m by Table \ref{tab.fitn_L},
the range 4.35--5.3 $\mu$m by Table \ref{tab.fitn_M},
the range 7.5--14.1 $\mu$m by Table \ref{tab.fitn_N},
and
the range 16--20 $\mu$m by Table \ref{tab.fitn_Q}\@.
The refractive index is chromatic (the phase shift depends on the 
wavelength $\lambda$ and wave number $\tilde\nu$) since 
\begin{equation}
dn/d\lambda= -\tilde\nu^2 \frac{dn}{d\tilde\nu}
\end{equation}
depends itself on $\lambda$.
(The negative of this parameter, measured in radian and divided by the areal
molar gas density, has been baptized
``normalized dispersion constant'' $K$ in \cite{MeisnerSPIE4838,MeisnerSPIE5491}.)
Our simple analytical fit format allows rapid calculation of
group refractive indices
as well \cite{MatharArxiv0605}.

The calculation adopts a standard of 370 ppmv of CO$_2$
as the most likely contemporary ambient clean air standard
\cite{globalview,YangGRL29,KaneJAtTP58}, well aware that laboratory
air may contain a higher volume fraction. Although adding this
mixing ratio as another free parameter to the procedure is feasible,
it has been kept fixed here to keep the size of the tables in check.

The use of $1/T$ rather than $T$ in the ansatz (\ref{eq.fit})
has no further relevance but aims to expand the validity of the
results to a large range of temperatures: in the simplest model
of the dispersion one expects the susceptibility to be proportional
to the molecular densities which are---with the ideal gas equation---proportional
to $p/T$.
This reasoning is actually void---as demonstrated by the
magnitudes of $c_{iTT}$---because we are employing non-infinite reference
temperatures $T_\mathrm{ref}$.

Solely for the benefit of the
reader who may use this type of result as a black box,
the parameter set in (\ref{eq.fit})
is based on relative humidity rather than some more fundamental measure
of the water molecule number density;
the more
appealing alternative from a scholarly point of view would have been
to split the computational steps into (i) equations to calculate absolute
molecular number densities, plus (ii) the fitting equations to transform these
to polarizabilities, and (iii) some post-processing with (\ref{eq.Lore2}). The
first step involves a self-consistent adaptation of the components
of the dry air at given mixing ratios to a partial pressure that is
``left over'' from $p$ after settling for the water density. The philosophy
behind Equation (\ref{eq.fit}) is to take this kind of burden away.

The negative values of $c_{0H}$ paraphrase that substitution of
the ``average'' dry air molecule by water at fixed total pressures $p$
decreases the
refractive index in all our wavelength
regions.

\begin{acknowledgments}
This work is supported by the NWO VICI grant of 15-6-2003
``Optical Interferometry: A new Method for Studies of Extrasolar Planets'' to A. Quirrenbach.
\end{acknowledgments}

\bibliographystyle{apsrmp}

\bibliography{all}

\end{document}